\def\tr{{\rm tr}\,}

\def\wt{\widetilde}
\def\sgn{{\rm sgn\,}}
\def\b{\bibitem}
\def\be{\begin{equation}}
\def\ee{\end{equation}}
\def\bea{\begin{eqnarray}}
\def\eea{\end{eqnarray}}
\def\bml{\begin{mathletters}}
\def\eml{\end{mathletters}}
\documentstyle[aps,prb,eqsecnum,psfig,epsf,floats]{revtex}
\draft
\begin{document}
\def\SNG{{\em Physical Review Style and Notation Guide}}
\def\LUG {{\em \LaTeX{} User's Guide \& Reference Manual}}
\def\btt#1{{\tt$\backslash$\string#1}}%
\def\REVTeX{REV\TeX}
\def\AmS{{\protect\the\textfont2
        A\kern-.1667em\lower.5ex\hbox{M}\kern-.125emS}}
\def\AmSLaTeX{\AmS-\LaTeX}
\def\BibTeX{\rm B{\sc ib}\TeX}
\twocolumn[\hsize\textwidth\columnwidth\hsize\csname@twocolumnfalse%
\endcsname

\title{Anderson-Mott Transition in a Magnetic Field:
        Corrections to Scaling
}
\author{D.Belitz}
\address{Department of Physics and Materials Science Institute\\
University of Oregon,\\
Eugene, OR 97403}
\author{T.R.Kirkpatrick}
\address{Institute for Physical Science and Technology, and Department of
 Physics\\
 University of Maryland,\\
 College Park, MD 20742}
\date{\today}
\maketitle

\begin{abstract}
It is shown that the Anderson-Mott metal-insulator transition of 
paramagnetic, interacting disordered electrons in an external magnetic 
field is in the same universality class as the transition from a 
ferromagnetic metal to a ferromagnetic insulator discussed recently. 
As a consequence, large corrections to scaling exist in the magnetic-field 
universality class, which have been neglected in previous theoretical 
descriptions. The nature and consequences of these corrections to scaling 
are discussed.
\end{abstract}
]

\section{Introduction}
\label{sec:I}

The metal-insulator transition of interacting, disordered
electrons at zero temperature ($T=0$), or Anderson-Mott transition (AMT),
has been studied theoretically near
two-dimensions ($d=2 + \epsilon$),\cite{F,F2,R} as well as in high dimensions
($d>6$,\cite{us_hi-d} and $d\alt 6$,\cite{us_glass}). 
In the latter case one finds the complete universality that is characteristic
of Landau theories, while in $d=2+\epsilon$ there is a large number of
different universality classes. This is because the AMT in low dimensions
is driven by soft modes, and the soft-mode spectrum of electrons at $T=0$
is determined by the symmetries of the underlying field theory. Accordingly,
systems in external magnetic fields, in the presence of magnetic impurities,
in the presence of spin-orbit scattering, etc., all constitute different 
universality classes. A priori it is not clear whether the high-dimensional
or the low-dimensional theories provide a better description of the situation
in $d=3$. Experimentally, while different universality classes
seem to exist, the situation is also far from clear.\cite{R} A serious problem 
is
posed by the fact that the minimum distance from the critical point, and the
ease and accuracy with which the distance from the critical point can be 
controlled, are much inferior to what can be routinely achieved near thermal
phase transitions. Consequently, one should expect corrections to scaling to
play an important role in the experimentally accessible region of parameter
space, and their incorporation into any interpretation of data to be 
essential. This expectation notwithstanding, however, 
corrections to scaling near an AMT have been largely
ignored by both experimentalists and theorists, with the exception of a
suggestion that strong (logarithmic) corrections to scaling may be responsible
for the ill-understood observations in phosphorus-doped silicon.\cite{us_logs,R}

Recently, another example of an AMT has been considered near two-dimensions, 
namely, the quantum
phase transition from a ferromagnetic metal to a ferromagnetic 
insulator.\cite{us_fm_mit} Since both an intrinsic magnetization and an
external magnetic field break the rotational symmetry in spin space and
give a mass to the transverse spin-triplet particle-hole excitations,
one expects this transition to be related to the AMT in an external field. 
An important question in this context relates to the Goldstone modes or spin 
waves that are present in a ferromagnet, but not in a system in an external 
field.
Since the Goldstone modes contribute to the soft-mode spectrum, one might
expect them to influence the critical behavior. However, it has been shown
in Ref.\ \onlinecite{us_fm_mit} that this is not the case. This makes it
likely that the two transitions indeed belong to the same universality
class. The effective model for the ferromagnetic MIT that was derived and 
studied 
in Ref.\ \onlinecite{us_fm_mit}, however, is different from the
one for the magnetic-field universality class considered before,\cite{F2}
and the former contains strong corrections to scaling that were absent in
the latter. This suggests that the existing description of the
magnetic-field universality class\cite{F2} is incomplete.

In this note we show that the two transitions belong indeed to the same
universality class, as has been suggested in Ref.\ \onlinecite{us_fm_mit},
and that the previously studied model for the magnetic-field AMT missed
terms that contribute important corrections to scaling. As a result of these
corrections to scaling, a reliable experimental determination of the critical
exponents is not possible from the currently available experimental data.

\section{Effective action}
\label{sec:II}

We consider the usual microscopic model for an AMT in an external magnetic 
field.\cite{F2,R} That is, we assume that the orbital effects of the magnetic
field serve only to suppress the soft modes in the particle-particle or
Cooper channel,\cite{orbital_footnote} and consider only the Zeeman term
explicitly. The latter's contribution to the action is
\bea
S_{\rm B}&=&\frac{1}{2}\,g_{\rm L}\,\mu_{\rm B}\,B\int dx\,T\sum_n\sum_{\alpha}
            \left[ {\bar\psi}_{n\uparrow}^{\alpha}({\bf x})\,\psi_{n\uparrow}
              ^{\alpha}({\bf x})  \right.
\nonumber\\
&&\hskip 50pt \left. - 
         {\bar\psi}_{n\downarrow}^{\alpha}({\bf x})\,\psi_{n\downarrow}
              ^{\alpha}({\bf x})\right]  \quad.
\label{eq:2.1}
\eea
Here $g_{\rm L}$ is the g-factor, $\mu_{\rm B}$ is the Bohr magneton,
$B$ is the magnetic field, and $T$ is the temperature. 
${\bar\psi}_{\uparrow,\downarrow}$ and 
$\psi_{n\uparrow,\downarrow}$ are fermionic fields for up- and down-spin 
electrons, with $n$ the index of a fermionic Matsubara frequency,
$\omega_n = 2\pi T (n+1/2)$, and $\alpha$ a
replica index to deal with the quenched disorder.

Within the formalism of Ref.\ \onlinecite{us_fm_mit}, bilinear products of
fermion fields are expressed in terms of a classical matrix field $Q$,
whose elements are related to the fermion fields by means of the isomorphism
\be
Q_{12} \cong \frac{i}{2}\,\left( \begin{array}{cccc}
          -\psi_{1\uparrow}{\bar\psi}_{2\uparrow} &
             -\psi_{1\uparrow}{\bar\psi}_{2\downarrow} &
                 -\psi_{1\uparrow}\psi_{2\downarrow} &
                      \ \ \psi_{1\uparrow}\psi_{2\uparrow}  \\
          -\psi_{1\downarrow}{\bar\psi}_{2\uparrow} &
             -\psi_{1\downarrow}{\bar\psi}_{2\downarrow} &
                 -\psi_{1\downarrow}\psi_{2\downarrow} &
                      \ \ \psi_{1\downarrow}\psi_{2\uparrow}  \\
          \ \ {\bar\psi}_{1\downarrow}{\bar\psi}_{2\uparrow} &
             \ \ {\bar\psi}_{1\downarrow}{\bar\psi}_{2\downarrow} &
                 \ \ {\bar\psi}_{1\downarrow}\psi_{2\downarrow} &
                      -{\bar\psi}_{1\downarrow}\psi_{2\uparrow} \\
          -{\bar\psi}_{1\uparrow}{\bar\psi}_{2\uparrow} &
             -{\bar\psi}_{1\uparrow}{\bar\psi}_{2\downarrow} &
                 -{\bar\psi}_{1\uparrow}\psi_{2\downarrow} &
                      \ \ {\bar\psi}_{1\uparrow}\psi_{2\uparrow} \\
                    \end{array}\right)
\label{eq:2.2}
\ee
Here all fields are understood to be taken at position ${\bf x}$, and
$1\equiv (n_1,\alpha_1)$, etc. In terms of these matrix fields, the
Zeeman term can be written\cite{i_footnote}
\be
{\cal A}_{\rm B} = -ih\int d{\bf x}\ \tr\left[(\tau_3\otimes s_3)\,
                   Q({\bf x}) \right]\quad,
\label{eq:2.3}
\ee
where $h=g_{\rm L}\,\mu_{\rm B}\,B$, 
$\tr$ is a trace over all discrete indices 
of the matrix $Q$, and $\tau_3 = -s_3 = -i\sigma_z$, with
$\sigma_z$ a Pauli matrix. More generally, it is useful to expand $Q$ into
a spin-quaternion basis,
\be
Q_{12}({\bf x}) = \sum_{r=0,3}\ \sum_{i=0,3} {^i_rQ}_{12}({\bf x})\,\left(
   \tau_r\otimes s_i\right)\quad,
\label{eq:2.4}
\ee
with $\tau_{0} = s_{0} = \openone$ the unit $2\times 2$ matrix.
The partition function can be written in terms of a functional integration
over the field $Q$ and an auxiliary field ${\wt\Lambda}$,\cite{us_fm_mit}
\bml
\label{eqs:2.5}
\be
Z = \int D[Q]\,D[{\wt\Lambda}]\ e^{{\cal A}[Q,{\wt\Lambda}]}\quad,
\label{eq:2.5a}
\ee
with an action
\be
{\cal A}[Q,{\wt\Lambda}] = {\cal A}_{{\rm B}=0}[Q,{\wt\Lambda}]
                           + {\cal A}_{\rm B}[Q]\quad.
\label{eq:2.5b}
\ee
\eml%
The action in the absence of a magnetic field, ${\cal A}_{{\rm B}=0}$, is
the same as in Ref.\ \onlinecite{us_fm_mit}.

The further formal development proceeds in exact analogy to 
Ref.\ \onlinecite{us_fm_mit}, and we will therefore be very brief.
The action allows for a saddle-point solution that can be written in
terms of two Green functions, ${\cal G}$ and ${\cal F}$, which are
related to the saddle-point values of the matrix elements ${^0_0Q}$ and
${^3_3Q}$, respectively. They obey the equations
\bml
\label{eqs:2.6}
\bea
\left(i\omega_n - \xi_{\bf k} - \Sigma_n\right)\,{\cal G}_n({\bf k})
        + \Delta_n\,{\cal F}_n({\bf k}) = 1\quad,
\label{eq:2.6a}\\
\left(i\omega_n - \xi_{\bf k} - \Sigma_n\right)\,{\cal F}_n({\bf k})
        + \Delta_n\,{\cal G}_n({\bf k}) = 0\quad.
\label{eq:2.6b}
\eea
\eml%
Here $\xi_{\bf k} = {\bf k}^2/2m - \epsilon_{\rm F}$ with $m$ the electron mass
and $\epsilon_{\rm F}$ the Fermi energy. The two self-energies $\Sigma$ 
and $\Delta$ are given by
\bml
\label{eqs:2.7}
\bea
\Sigma_n &=& \frac{1}{\pi N_{\rm F}\tau}\,\frac{1}{V}\sum_{\bf k}
           {\cal G}_n({\bf k})\quad,
\label{eq:2.7a}\\
\Delta_n &=& \Delta - \frac{1}{\pi N_{\rm F}\tau}\,\frac{1}{V}\sum_{\bf k}
           {\cal F}_n({\bf k})\quad,
\label{eq:2.7b}
\eea
with
\be
\Delta = h + 2\Gamma^{(t)}T\sum_n\frac{1}{V}\sum_{\bf k}{\cal F}_n({\bf k})
         \quad.
\label{eq:2.7c}
\ee
\eml%
Here $\Gamma^{(t)}$ is the spin-triplet interaction amplitude defined in
Ref.\ \onlinecite{us_fm_mit}, $\tau$ is the transport relaxation time
due to the quenched disorder, and $N_{\rm F}$ is the density of states
at the Fermi level. The appearance of $h$ in Eq.\ (\ref{eq:2.7c})
is the only difference between the saddle-point equations here and those
in Ref.\ \onlinecite{us_fm_mit}, and it results from the Zeeman term in
the action. For later reference we note that the field configuration obtained
by putting $\Delta = 0$ is {\em not} a saddle point, in constrast to the
situation in the absence of an external field.

The discussion of the saddle-point solution proceeds as in
Ref.\ \onlinecite{us_fm_mit}. In particular, $\Delta$ obeys the equation
\bml
\label{eqs:2.8}
\be
\Delta = h - T\sum_n\frac{1}{V}\sum_{\bf k}
         \frac{2\Gamma^{(t)}\Delta}
     {(i\omega_n - \xi_{\bf k} + i\sgn\omega_n/2\tau)^2 - \Delta^2}\quad.
\label{eq:2.8a}
\ee
For $h=0$ there is a nonzero physical solution only if the Stoner criterion
$N_{\rm F}\Gamma^{(t)} > 1$ is fulfilled.\cite{us_fm_mit} In paramagnetic
systems, $N_{\rm F}\Gamma^{(t)}$ is smaller than the critical value, and
$\Delta=0$ in the absence of a magnetic field. However, for
nonzero $h$ we have
\be
\Delta = h\,\left[1 + N_{\rm F}\Gamma^{(t)}\right] + O(b^3)\quad.
\label{eq:2.8b}
\ee
\eml%
Since $\Delta$ is proportional to the magnetization,\cite{us_fm_mit} this
reflects the fact that a magnetic field polarizes the spins.

We see that, at the saddle-point level, the only difference between the
current situation and the one considered in Ref.\ \onlinecite{us_fm_mit}
is the origin of the nonzero value of $\Delta$. More generally, as far as an
effective theory that concentrates on soft-mode effects is concerned, the
only difference lies in the existence of Goldstone modes in a magnetic
system. Since the latter turn out to be irrelevant for describing the
metal-insulator transition,\cite{us_fm_mit} the procedure of expanding
about the saddle-point, separating soft and massive modes, and deriving an
effective theory for the metal-insulator transition is therefore the
same in the magnetic field and the ferromagnetic cases, respectively, 
and we can simply take over the
result for the latter. The final effective action thus reads\cite{us_fm_mit}
\bml
\label{eqs:2.9}
\bea
{\cal A} &=& \frac{-1}{2G}\int d{\bf x}\ \tr[\nabla{\wt Q}({\bf x})]^2
                + 2H\int d{\bf x}\ \tr [\Omega\,{\wt Q}({\bf x})]
\nonumber\\
&&\hskip -7pt - \frac{1}{2G_3}\int d{\bf x}\ \tr\left((\tau_3\otimes s_3)
         [\nabla{\wt Q}({\bf x})]^2\right)
\nonumber\\
&&\hskip -7pt + 2H_3\int d{\bf x}\ \tr\left[(\tau_3\otimes s_3)\,\Omega\,
   {\wt Q}({\bf x}) \right] + {\cal A}_{\rm int}[{\wt Q}]\ .
\label{eq:2.9a}
\eea
Here
\be
\Omega_{12} = (\tau_0\otimes s_0)\,\delta_{12}\,2\pi Tn\quad,
\label{eq:2.9b}
\ee
is a bosonic Matsubara frequency matrix. The bare values of the
coupling constants $G$, $G_3$, $H$, and $H_3$ are
\bea
1/G&=&\frac{\pi}{4}\,m\,(\sigma_0^+ + \sigma_0^-)\quad,
\label{eq:2.9c}\\
1/G_3&=&\frac{\pi}{4}\,m\,(\sigma_0^+ - \sigma_0^-)\quad,
\label{eq:2.9d}\\
H&=&\frac{\pi}{16}\,(N_{\rm F}^+ + N_{\rm F}^-)\quad,
\label{eq:2.9e}\\
H_3&=&\frac{\pi}{16}\,(N_{\rm F}^+ - N_{\rm F}^-)\quad,
\label{eq:2.9f}
\eea
where $\sigma_0^{\pm}$ and $N_{\rm F}^{\pm}$ are the Boltzmann conductivity
and the bare density of states at the Fermi level, respectively, of systems
whose Fermi energy has been shifted by $\pm\Delta$ from its level in the
absence of a magnetic field. Notice that $1/G_3$ and $H_3$ vanish for a
vanishing magnetic field, and for small fields are linear in the field,
see Eqs.\ (\ref{eq:2.9d},\ref{eq:2.9f},\ref{eq:2.8b}).
The interaction piece of the action consists of three terms,
\be
{\cal A}_{\rm int}[Q] = {\cal A}_{\rm int}^{(s)}[Q]
                         + {\cal A}_{\rm int}^{(t)}[Q]
                         + {\cal A}_{\rm int}^{(3)}[Q]\quad,
\label{eq:2.9g}
\ee
\bea
{\cal A}_{\rm int}^{(s)}[Q]&=& \frac{-\pi T}{4}\,K_s\int d{\bf x}\ \sum_{1234}
   \delta_{\alpha_1\alpha_2}\,\delta_{\alpha_1\alpha_3}\,\delta_{1-2,4-3}\,
\nonumber\\
\hskip -20pt &&\times\sum_r (-)^r \tr\left[(\tau_r\otimes s_0)\,Q_{12}({\bf x})
   \right]\
\nonumber\\
\hskip -20pt &&\times\tr\left[(\tau_r\otimes s_0)\,Q_{34}({\bf x})\right]\quad,
\label{eq:2.9h}
\eea
\bea
{\cal A}_{\rm int}^{(t)}[Q]&=& \frac{-\pi T}{4}\,K_t\int d{\bf x}\ \sum_{1234}
   \delta_{\alpha_1\alpha_2}\,\delta_{\alpha_1\alpha_3}\,\delta_{1-2,4-3}\,
\nonumber\\
&&\hskip -20pt\times\sum_r (-)^r \tr\left[(\tau_r\otimes s_3)\,Q_{12}({\bf x})
    \right]\
\nonumber\\
 &&\hskip -20pt\times\tr\left[(\tau_r\otimes s_3)\,Q_{34}({\bf x})\right]\quad,
\label{eq:2.9i}
\eea
\bea
{\cal A}_{\rm int}^{(3)}[Q]&=& -4\pi TK_3\int d{\bf x}\ \sum_{1234}
   \delta_{\alpha_1\alpha_2}\,\delta_{\alpha_1\alpha_3}\,\delta_{1-2,4-3}\,
\nonumber\\
&& \sum_{rs}\sum_{ij} m_{rs,ij}\ {^i_r Q}_{12}({\bf x})\,{^j_s Q}_{34}({\bf x})
   \quad,
\label{eq:2.9j}
\eea
where
\be
m_{rs,ij} = \frac{1}{4}\,\tr (\tau_3\tau_r\tau_s^{\dagger})\
                         \tr (s_3 s_i s_j^{\dagger})\quad,
\label{eq:2.9k}
\ee
and $K_t = 2\pi\Gamma^{(t)}$ and $K_s$ are the usual spin-triplet and
spin-singlet interaction amplitudes that are also present in the absence
of a magnetic field.\cite{R} $K_3$, like $1/G_3$ and $H_3$, is magnetic
field dependent, and vanishes linearly with the field for small fields.
Since it is absent from the bare action, but generated by the renormalization 
group at one-loop order,\cite{us_fm_mit} it is also proportional to the 
disorder. Finally,
\be
{\wt Q}_{12} = {\hat Q}_{12} - \delta_{12}\,(\tau_0\otimes s_0)\,\omega_{n_1}
               \quad,
\label{eq:2.9l}
\ee 
and ${\hat Q}$ is subject to the constraints
\be
{\hat Q}^2({\bf x}) \equiv 1\quad,\quad
{\hat Q}^{\dagger} = {\hat Q}\quad,\quad
\tr{\hat Q}({\bf x}) \equiv 0\quad.
\label{eq:2.9m}
\ee
\eml%

\section{Metal-insulator transition}
\label{sec:III}

We recognize Eqs.\ (\ref{eqs:2.9}) as the generalized nonlinear $\sigma$
model proposed by Finkel'stein\cite{F2} for the magnetic field universality
class, except that the latter was missing the terms proportional to
$1/G_3$, $H_3$, and $K_3$. As was shown in Ref.\ \onlinecite{us_fm_mit},
these terms do not change the asymptotic critical behavior. Choosing the
correlation length exponent
$\nu$, the critical exponent for the density of states $\beta$, and
the dynamical critical exponent $z$ as the three independent exponents,
we thus have, to lowest order in an expansion in 
$\epsilon = d-2$,\cite{F2,R,LR_footnote}
\bml
\label{eqs:3.1}
\bea
\nu&=&1/\epsilon + O(1)\quad,
\label{eq:3.1a}\\
\beta&=&1/2\epsilon (1 - \ln 2)\quad,
\label{eq:3.1b}\\
z&=&d \quad,
\label{eq:3.1c}
\eea
and the critical exponent for the conductivity, $s=\nu (d-2)$, is
\be
s = 1 + O(\epsilon)\quad.
\label{eq:3.1d}
\ee
\eml%

While the additional coupling constants $1/G_3$, $H_3$, and $K_3$ do
not contribute to the leading critical behavior, they lead to corrections
to scaling. As was shown in Ref.\ \onlinecite{us_fm_mit}, the least
irrelevant operator related to these coupling constants has a scale
dimension
\be
\lambda_3 = -\,\frac{3\ln 2 - 2}{1 - \ln 2}\,\epsilon + O(\epsilon^2)\quad.
\label{eq:3.2}
\ee
Consequently, in the vicinity of the metal-insulator transition any
observable $\omega$ obeys the homogeneity law
\be
\omega(t,T,u,\ldots) = b^{-x_\omega}\omega(tb^{1/\nu},Tb^z,ub^{\lambda_3},
                       \ldots) \quad,
\label{eq:3.3}
\ee
where $t$ is the dimensionless distance from the critical point. Apart from
(or instead of) the temperature $T$, other generalized external fields may
appear as arguments of $\omega$, e.g., a frequency, or an electric field,
depending on the nature of the observable under consideration. $u$ denotes 
the least irrelevant variable (which is a linear
combination of $1/G_3$, $H_3$, and $K_3$),
$x_\omega$ is the scale dimension of $\omega$, 
$b$ is an arbitrary scale factor,
and the ellipses denote the dependence of $\omega$ on variables that are more
irrelevant than $u$. These corrections to scaling were absent from the
original model of Ref.\ \onlinecite{F2}.

\section{Discussion}
\label{sec:IV}

To put our results in context, we give a brief discussion of corrections to 
scaling in general, the additional terms in Eq.\ (\ref{eq:2.9a}) in particular, and
why they were missed in previous treatments. Asymptotically close to a critical
point, observables obey the homogeneity law, Eq.\ (\ref{eq:3.3}), and of
the various coupling constants only those with positive scale dimensions need
to be kept. Apart from the dimensionless distance from the critical point,
whose scale dimension is by definition the inverse correlation length
exponent $1/\nu >0$, the only coupling constants with positive scale dimensions 
are generalized external fields,\cite{MEF} like the
temperature in our example. Technically, the homogeneity law appears as the
solution of a renormalization group equation, which takes the form of a
Callan-Symanzik equation  or a similar partial differential equation. 
If one is not asymptotically close to the transition, corrections to the 
asymptotic scaling behavior appear from two sources:\cite{ZJ} (1) The solution 
of the renormalization group equation becomes more complicated than a 
generalized homogeneous function, and needs to be expanded in a power series.
(2) The ``irrelevant operators'', i.e. coupling constants whose scale 
dimensions are negative, can no longer be ignored.\cite{DIV_footnote}
Keeping them, and ordering them with respect to their irrelevancy, leads to the 
Wegner expansion.\cite{Wegner,ZJ} In our case, the
leading term in the Wegner expansion, which we have kept in Eq.\ (\ref{eq:4.1}),
is more important than the corrections to scaling of type (1). In fact, our 
value
$-\nu\lambda_3 \approx 0.26$ is rather small for a leading Wegner exponent, 
which
typically is on the order of 0.5. Since these large corrections to scaling 
result
from the coupling constants $1/G_3$, $H_3$, and $K_3$ in Eqs.\ (\ref{eqs:2.9}),
it is important to keep these terms.

This brings us to our next discussion topic, viz. the origin of these additional
terms in the action. As we have mentioned in Sec.\ \ref{sec:II} above, a field
configuration that has $\Delta = 0$, and thus describes non-spin polarized 
electrons, is not a saddle-point
solution of the field theory, Eqs.\ (\ref{eqs:2.5}), underlying the effective
action. Reference\ \onlinecite{F2} used the implicit assumption that the 
{\em only}
effect of the magnetic field, as far as the metal-insulator transition is
concerned, is to give a mass to certain modes that are massless in the absence
of an external field.\cite{R} If this were true, then one could use a field
configuration with $\Delta = 0$ as the starting
point for deriving an effective action. The result of this procedure\cite{F2} is
Eqs.\ (\ref{eqs:2.9}) with $1/G_3 = H_3 = K_3 = 0$. By the nature of the
renormalization group flow equations,\cite{us_fm_mit} zero bare values for these
coupling constants imply that they are not generated under renormalization 
either.
The effective theory that was originally proposed for this problem is therefore
incomplete. As we have seen, this omission has an influence, not on
the asymptotic scaling properties, but on the corrections to scaling, within the
resulting effective field theory. Since the additional terms discussed above
describe a physical effect, namely the
spin polarization due to the magnetic field that was neglected in the original
treatment, we conclude that the leading corrections to scaling discussed in this
paper are an immediate, if a priori not entirely obvious, consequence of this 
spin polarization. We also note that there is no obvious reason underlying our
result that the additional terms do not change the asypmtotic critical behavior;
this appears to be accidental. In particular, it needs to be stressed that
this conclusion holds only in an $\epsilon$-expansion near two-dimensions, and
in $d=3$ the new terms might well influence the asymptotic scaling behavior.

We finally discuss the experimental implications of our results. For 
definiteness, let us consider
the conductivity $\sigma$, whose scale dimension is $x_\sigma = s/\nu$.
A common way to analyze experiments is to extrapolate the data to $T=0$
and consider $\sigma(t,T=0)$. From Eq.\ (\ref{eq:3.3}), we have
\be
\sigma(t,T=0) = \sigma_0\,t^s\,\left[1 + {\rm const.}\times t^{-\nu\lambda_3}
                \right]\quad.
\label{eq:4.1}
\ee
Here $\sigma_0$ is a microscopic conductivity scale on the order of the
Boltzmann conductivity, and the constant is nonuniversal. Assuming that
the constant is of $O(1)$, in order to achieve, e.g., a 10\% accuracy
in a determination of the value of $s$ from a log-log plot, the $t$-range
needs to be restricted to 
$t < (0.1/\nu\vert\lambda_3\vert)^{1/\nu\vert\lambda_3\vert}$.
Extrapolating the results of our one-loop approximation to $d=3$, we have
$\nu\vert\lambda_3\vert \approx 0.26$, or $t < 0.026$. If the constant
is larger, as is the case at many critical points,\cite{Sengers} the
asymptotic critical region is even smaller. For instance, for 
${\rm const.} = 1/\vert\nu\lambda_3\vert \approx 4$, we need 
$t\alt 10^{-4}$ to measure $s$ with a 10\% accuracy.
Such small values of $t$ are not currently achievable for Anderson-Mott
transitions. The technique that yields the best control over $t$,
viz. stress tuning, allows to probe the region 
$10^{-3}\alt t\alt 10^{-2}$, while other methods are
restricted to $t\alt 10^{-2}$ or larger.\cite{R} 
We conclude that it is currently
not possible to determine the asymptotic critical exponents for the
magnetic-field universality class of the Anderson-Mott transition with
any meaningful accuracy.

\acknowledgments
Parts of this work were performed at the Institute for Theoretical Physics
at UC Santa Barbara, and at the Aspen Center for Physics. This work was 
supported by the NSF under grant Nos. DMR-98-70597, DMR-99-75259, 
and PHY94-07194.

\end{document}